\documentclass[12pt]{article}
\usepackage{amssymb}
\usepackage{graphicx}
\usepackage{layout}
\begin{document}
\title{Testing angular momentum effects on the space time}
\author{A. Tartaglia, M. L. Ruggiero \\
Dip. Fisica, Politecnico, and INFN, Torino, Italy\\ e-mail: tartaglia@polito.it ; ruggierom@polito.it}

\maketitle
\begin{abstract}
The paper contains a proposed experiment for testing the angular momentum effect on the propagation of light around
a rotating mass. The idea is to use a rotating spherical laboratory-scale shell, around which two mutually
orthogonal light guides are wound acting as the arms of an interferometer. Numerical estimates show that time of
flight differences between the equatorial and polar guides could be in the order of $\sim 10^{-20}$ s per loop.
Using a few thousands loops the time difference is brought in the range of feasible interference measurements.
\end{abstract}

\section{Introduction}

The general relativistic effects of rotating masses have since been
considered in connection with gravitomagnetism, i.e. with the part of the
gravitational field which displays a solenoidal character similar to that of
the magnetic field \cite{gravimag}. These effects are usually much less
relevant than those of the gravitoelectric (radial) part of the field.
Gravitomagnetism is expected to influence the precession of orbiting
gyroscopes (Lense-Thirring effect \cite{lense}), the synchronization of
clocks in the field of rotating masses \cite{mashhoon}, \cite{tartaglia},
the time of flight of light around spinning bodies \cite{tartaglia2}.

Evidence of gravitomagnetic effects are found studying the behavior of the
binary pulsar 1913+16 \cite{cliff} and considering the lunar orbit as
obtained by laser ranging, during the motion of the Earth-Moon pair in the
solar system\cite{cliff}. In general however the relevance of
gravitomagnetic effects in the field of the Earth is extremely small and the
only attempts to detect them have since been limited to the quoted lunar
orbit study, to the precession of the nodes of the orbits of LAGEOS\
satellites \cite{ciufolini} and to the precession of gyroscopes carried by a
spacecraft \cite{gpb}.

Gravitomagnetic effects originate, in weak field approximation, from the
extra-diagonal term of the metric in the vicinity of a rotating body. It
turns out however that the rotation of a body affects the diagonal terms too
introducing corrections, which, when the mass terms become so small to be
negligible, though small, remain nonetheless perceivable. Here we propose a
ground based experiment exploiting the effect on the time of flight of light
rays, induced by a rotating mass. Actually, as we shall see, the time of
flight is influenced both by the very mass $M$ of the central body and by
its angular momentum density, expressed by the parameter $a=J/(Mc)$ i.e. the
ratio between the angular momentum and the product of the mass by the speed
of light. However when considering the time of flight difference between an
equatorial and a polar circular trajectory, it turns out, at the lowest
significant order, to be proportional to $a^{2}$.

The actual value of $a$ depends on the geometry of the source and its
angular velocity. In this respect thin spherical shells perform better than
solid spheres. For the whole Earth $a$ is in the order of 4 m; at the
laboratory scale it is some orders of magnitude lower, however we shall show
that the final value for the time difference is within the sensitivity range
of interferometric techniques currently available. The expected cost for the
proposed on Earth experiment should in turn be much lower than those in
space.

\section{The time difference}

In an axially symmetric stationary gravitational field the null interval is
written:
\[
0=g_{tt}dt^{2}+2g_{t\phi }dtd\phi +g_{rr}dr^{2}+g_{\theta \theta }d\theta
^{2}+g_{\phi \phi }d\phi ^{2}
\]
where the metric elements do not contain either $t$ or $\phi $. Considering
a circular path the (coordinate) time of flight for an equatorial revolution
($\theta =\pi /2$) is \cite{tartaglia3}

\begin{equation}
T_{e}=2\pi \frac{\mp g_{t\phi }+\sqrt{g_{t\phi }^{2}-g_{tt}g_{\phi \phi }}}{%
g_{tt}}  \label{equatoriale}
\end{equation}
The $-$ sign stands for co-rotation, $+$ means counter-rotation.

Using Boyer-Lundquist coordinates in a Kerr metric (\ref{equatoriale})
becomes
\begin{equation}
T_{e}=\frac{2\pi }{c^{2}}\frac{\mp \frac{2GMa}{cr}+c\sqrt{\left( \frac{2GMa}{%
c^{2}r}\right) ^{2}+\left( 1-\frac{2GM}{c^{2}r}\right) \left( r^{2}+a^{2}+%
\frac{2GMa^{2}}{c^{2}r}\right) }}{1-\frac{2GM}{c^{2}r}}  \label{kerr}
\end{equation}

The other configuration we are considering is a fixed azimuth polar circular
trajectory. Now it is
\[
T_{p}=\int_{o}^{2\pi }\sqrt{-\frac{g_{\theta \theta }}{g_{tt}}}d\theta
\]
or explicitly
\begin{equation}
T_{p}=\frac{1}{c}\int_{o}^{2\pi }\sqrt{\frac{r^{2}+a^{2}\cos ^{2}\theta }{1-%
\frac{2\frac{GM}{c^{2}}r}{r^{2}+a^{2}\cos ^{2}\theta }}}d\theta
\label{polarkerr}
\end{equation}

The Kerr metric has been used both as an example of a well known axially
symmetric stationary metric and because its weak field limit coincides with
the result which can be obtained in a post Newtonian approximation applied
to the exterior of a spherical distribution of mass and energy \cite
{straumann}.

In a weak field and introducing the small parameters $\mu =\frac{2GM}{c^{2}r}
$ and $\alpha =a/r$ (\ref{kerr}) and (\ref{polarkerr}) become
\begin{eqnarray}
T_{e} &=&2\pi \sqrt{-\frac{g_{\phi \phi }}{g_{tt}}}=\frac{2\pi }{c}R\left( 1+%
\frac{1}{2}\alpha ^{2}+\allowbreak \frac{1}{2}\mu \right)  \nonumber \\
T_{p} &=&\int_{o}^{2\pi }\sqrt{-\frac{g_{\theta \theta }}{g_{tt}}}d\theta =%
\frac{R}{c}\int_{0}^{2\pi }\left( \allowbreak 1+\frac{1}{2}\alpha ^{2}\cos
^{2}\theta +\frac{1}{2}\mu \right) d\theta  \label{tempi} \\
&=&\allowbreak 2\pi \frac{R}{c}\left( 1+\frac{1}{2}\mu +\frac{1}{4}\alpha
^{2}\right)  \nonumber
\end{eqnarray}
In $T_{e}$ the term corresponding to the first one on the numerator of eq.(%
\ref{kerr}) has been overlooked because at the laboratory scale where we
imagine to be working the product $\mu a$ is many orders of magnitude
smaller than $a^{2}$.

In practice, at the lowest approximation order, the only difference between $%
g_{\phi \phi }$ and $g_{\theta \theta }$ is in a factor $\cos ^{2}\theta $
multiplying the $\alpha ^{2}$ correction; that factor is responsible,
through the integration, for the $1/4$ factor appearing in front of $\alpha
^{2}$ in $T_{p}$.
\begin{equation}
\Delta T=T_{e}-T_{p}=\allowbreak \frac{1}{2}\frac{\pi }{cR}a^{2}
\label{differenza}
\end{equation}
which, as can be seen, depends only on $a^{2}$ (as said, the terms
containing the mass mix it with $a$ and are smaller).

Of course (\ref{differenza}) corresponds to the hypothesis of two exactly
equal radii circumferences. This perfect equality would in practice be
impossible to obtain, so let us rewrite the result considering two different
radii $R$ $(\equiv R_{p})$ and $R_{e}=R_{p}+\delta R$. One has (lowest order
in $\delta R/R$)
\begin{equation}
\Delta T=2\pi \frac{\delta R}{c}+\frac{\pi }{2}\frac{a^{2}}{cR}\left( 1-%
\frac{\delta R}{R}\right)  \label{disuguali}
\end{equation}

\section{Laboratory scale}

For a homogeneous steadily rotating sphere it is
\begin{equation}
a_{f}=\frac{2}{5}\frac{R^{2}}{c}\Omega  \label{pianeti}
\end{equation}
where $\Omega $ is the angular speed of the sphere.

The value of $a$ can be increased a bit considering instead of a sphere a
hollow spherical thin shell. In that case one has
\begin{equation}
a_{h}=\frac{2}{3}\frac{R^{2}}{c}\Omega  \label{guscio}
\end{equation}
Now however the mass, for the same external radius, is much lower than
before. Actually
\[
\frac{M_{h}}{M_{f}}=\allowbreak 3\frac{h}{R}
\]
where $h$ is the thickness of the shell ($h<<R$).

Applying (\ref{pianeti}) to the Earth the result is
\[
a_{E}=3.9\textit{ m}
\]

Other examples are Jupiter or the Sun \cite{tartcqg17}:
\begin{eqnarray*}
a_{J} &=&1.2\times 10^{3}\textit{ m} \\
a_{S} &\simeq &3.0\times 10^{3}\textit{ m}
\end{eqnarray*}

In the laboratory one can of course expect much lower values. Let us
consider a hollow sphere as the source of the effect. The $a$ value is
limited in practice by the strength of the wall of the shell. In fact the
resulting centrifugal force on a hemisphere is
\[
F_{c}\allowbreak =\pi \rho h\Omega ^{2}R^{3}
\]
($\rho $ is the density of the material).

The corresponding average tension induced in the wall of the shell is
\[
<\sigma >=\frac{\pi \rho h\Omega ^{2}R^{3}}{2\pi Rh}=\allowbreak \frac{1}{2}%
\rho \Omega ^{2}R^{2}
\]
The maximum stress is attained at the equator, being in the order
(unidimensional stresses are assumed):
\[
\sigma _{m}=\allowbreak \rho \Omega ^{2}R^{2}
\]

If $\sigma _{m}$ coincides with the allowable resistance of the material the
maximal peripheral velocity is $v_{m}=\sqrt{\sigma _{m}/\rho }$. The
attainable value of $a_{h}$ can consequently be written in terms of the
properties of the material: $a_{h}=\frac{2}{3}\frac{R}{c}\sqrt{\sigma
_{m}/\rho }$.

Finally the $a$-dependent part of the time difference (\ref{differenza})
becomes
\begin{equation}
\Delta T=\allowbreak \frac{2}{9}\frac{\pi }{c^{3}}\frac{R^{2}}{R_{l}}\frac{%
\sigma _{m}}{\rho }  \label{pratico}
\end{equation}
Here $R_{l}$ is the radius of the light's path (a little bit greater than $R$%
).

Considering composite materials $\sigma _{m}$ can be as high as 2000 MPa,
with a density $\rho \sim 1700$ kg/m$^{3}$ \cite{romeo}. These values lead
to (assuming, just to fix ideas, $R=1$ m)
\begin{eqnarray*}
a_{h} &=&2.4\times 10^{-6}\textit{ m} \\
\Delta T &=&3\times 10^{-20}\textit{ s}
\end{eqnarray*}
Using visible light the relative phase shift corresponding to the time of
flight difference is in the order of $10^{-5}$.

\section{Proposing an actual experiment}

A phase shift like the one computed in the previous section is of the same
order of magnitude as the phase differences expected in gravitational waves
interferometric detection experiments \cite{virgo}. The advantage now could
be the comparatively small size of the apparatus. The idea is to have as a
gravitomagnetic source a spinning thin spherical shell; to fix a hypothesis
it could have a $1$ m radius and a wall thickness of $1$ mm or less. Such
object would weigh not more than 209 N and should rotate at a maximum
angular speed of $\Omega _{m}\simeq 10^{3}$ rad/s. Two circular non rotating
light guides should contour the sphere, one at the equator, the other
through the poles, as in figure 1. A laser beam would be split at $A$, the
two resulting secondary beams would be guided along the two circular paths
and finally would be led to interfere at $B$.
\begin{figure}[top]
\begin{center}
\includegraphics[width=9cm,height=7cm]{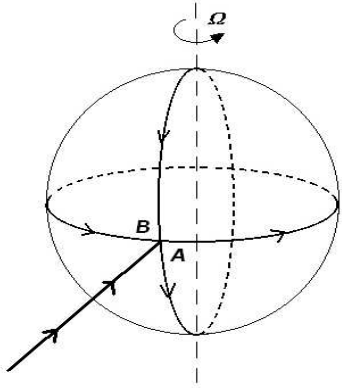}
\caption{ The hollow sphere rotates at the angular speed $\Omega $%
. The two circular wave guides are fixed. A primary light beam is split in $%
A $ to follow separately the equatorial and the polar trajectories, then the beam is recombined interfering in
$B.$} \label{fig:fig1}
\end{center}
\end{figure}

The beam intensity relative change at the interference $\delta I/I$ is related to the phase shift $\delta \Phi $
according to
\begin{equation}
\frac{\delta I}{I}=\frac{1}{2}\left( 1-\cos \delta \Phi \right)
\label{intensity}
\end{equation}
Using the estimates of the previous section, this means (for the
contribution depending on $a^{2}$)
\[
\frac{\delta I}{I}\sim 10^{-9}
\]

It is not possible to extract such an intensity fluctuation if it is static;
furthermore the inevitable difference in the lengths of the two optical
fibers must be taken into account. This means that we need modulating in
time the effect we are looking for; this result can be achieved in principle
periodically varying the angular speed of the shell. However a better
solution would be to keep the angular velocity constant but to let the
rotation axis precess about a direction half way between the two orthogonal
planes containing the optical fibers loops. In this way the role of the two
fibers (the 'arms' of the interferometer) would be periodically interchanged
with a frequency double the precession frequency.

Actually the time of flight difference is a cumulative effect. This means
that one can think of winding for instance $10^{4}$ loops of an optical
fiber. This would correspond to a total length in the order of 63 km with a
total thickness of the coil in the order of 1 cm (continuous fibers that
long and as thin as 1 $\mu $m in diameter are presently available). The
final time difference would then be
\begin{equation}
\Delta T_{t}\sim 10^{-16}\textit{ s}  \label{diftot}
\end{equation}
In the case of visible light this difference corresponds to $1/10$ of a
period or, in terms of interference patterns, to $1/10$ of a fringe. The
attainable intensity modulation, from (\ref{intensity}), would be
\[
\frac{\delta I}{I}\sim 10^{-1}
\]

\section{Conclusion}

We have shown that it might be possible to realize a ground based experiment
to reveal rotation effects on the structure of space time using a laboratory
size rotating mass. In fact available materials (composite carbon fibers
high resistance materials) and available technologies for detection of very
small periodically varying intensity perturbations in a light signal do
allow for the possibility to measure the time difference (\ref{differenza})
and even more (\ref{diftot}) and consequently the influence of the angular
momentum density around a spinning body. Different radii of the fibers lead
to a term (the first one in (\ref{disuguali})) not depending on the angular
speed of the source (i.e. on $a$) and to a correction of the $a^{2}$
dependent term proportional to $\delta R/R$. The $a$-independent term is
neutralized modulating in time the signal via the precession of the rotation
axis of the source. The $\delta R/R$ correction can easily be in the order
of one part in $10^{4}$ (let us say 0.1 mm compared to 1 m) or lower, which
we can consider as marginal.

Of course a lot of technical details require study and consideration, but
with no higher difficulty than the problems implied by interferometric
detection of gravitational waves. The advantage of our proposal would be to
have a cost presumably much lower than other experiments aimed to verify
weak general relativistic effects.

\subsection{Acknowledgment}

The authors would like to thank Ron Adler for suggesting the idea of the
precession of the rotation axis of the hollow sphere as a means to modulate
the interference pattern in time.

\end{document}